# Perspective on nanoscale magnetic sensors using giant anomalous Hall effect in topological magnetic materials for read head application in magnetic recording

Tomoya Nakatani,[*] Prabhanjan D. Kulkarni,[**] Hirofumi Suto, Keisuke Masuda, Hitoshi Iwasaki, and Yuya Sakuraba

*Research Center for Magnetic and Spintronic Materials, National Institute for Materials Science, 1-2-1 Sengen, Tsukuba, Ibaraki, 305-0047, Japan*

[*] Corresponding author. Email: nakatani.tomoya@nims.go.jp
[**] Present address: SEMITEC Corporation, Chiba, Japan

**Abstract**

Recent advances in the study of materials with topological electronic band structures have revealed magnetic materials exhibiting giant anomalous Hall effects (AHE). The giant AHE has not only attracted the research interest in its mechanism but also opened up the possibility of practical application in magnetic sensors. In this article, we describe simulation-based investigations of AHE magnetic sensors for the applications to read head sensors (readers) of hard disk drives. With the shrinking of magnetic recording patterns, the reader technology, which currently uses multilayer-based tunnel magnetoresistance (TMR) devices, is associated with fundamental challenges, such as insufficient spatial resolution and signal-to-noise ratio (SNR) in sensors with dimensions below 20 nm. The structure of an AHE-based device composed of a single ferromagnetic material is advantageous for magnetic sensors with nanoscale dimensions. We found that AHE readers using topological ferromagnets with giant AHE, such as $Co_2MnGa$, can achieve a higher SNR than current TMR readers. The higher SNR originates from the large output signal of the giant AHE as well as from the reduced thermal magnetic noise, which is the dominant noise in TMR readers. We highlight a major challenge in the development of AHE readers: the reduction in the output signal due to the shunting of the bias current and the leakage of the Hall voltage through the soft magnetic shields surrounding the AHE reader. We propose reader structures that overcome this challenge. Finally, we discuss the scope for future research to realize AHE readers.

## I. Introduction: Anomalous Hall effect, materials, and magnetic sensor

The ordinary Hall effect (OHE) occurs when a voltage potential difference is created across an electrical conductor by the Lorentz force acting in the transverse direction to the electric current and magnetic field. This effect has been utilized in magnetic sensors for various applications, such as in the detection of geomagnetic fields, electric currents, rotation, and speed of moving objects.[1–3] The magnitude of the OHE is expressed using the ordinary Hall coefficient, $R_0 = \frac{1}{ne}$, where we assume that one type of charge carrier (electron or hole) dominates the electronic conduction of the material, and $n$ and $e$ denote the carrier density and the charge of the carrier, respectively. In the free electron model, $R_0 = \rho\mu$, where $\rho$ ($= \frac{m^*}{ne^2\tau}$) and $\mu$ ($= \frac{e\tau}{m^*}$) are the resistivity and the carrier mobility, respectively ($m^*$: effective mass, and $\tau$: relaxation time). Thus, $n$-type semiconductors with a high electron mobility, such as GaAs, InAs, and InSb, are used as OHE-based magnetic sensors. In addition to the OHE, conductive magnetic materials, including ferromagnets, ferrimagnets, and limited types of antiferromagnets, exhibit another type of Hall effect; the anomalous Hall effect (AHE). Figures 1(a) and (b) show the schematic Hall cross of a ferromagnetic thin film and the Hall voltage ($V_H$) vs. out-of-plane magnetic field ($H_z$) curve, respectively. The ferromagnetic film is in-plane magnetized when $H_z = 0$. $V_H$ is proportional to the out-of-plane component of the magnetization of the ferromagnet ($M_z$). The saturation magnetic field ($H_s$) is determined by the out-of-plane demagnetizing field, $\mu_0 M_s$, in the case of ferromagnets without perpendicular magnetic anisotropy, where $M_s$ is the saturation magnetization. Since the $V_H$–$H_z$ curves show a good linearity for $H_z < H_s$, AHE devices are also suitable for magnetic field sensors. The Hall resistivity ($\rho_H$) can be empirically expressed as $\rho_H = -\rho_{xy} = R_0 \mu_0 H_z + R_s M_z$, where $\rho_{xy}$ is the off-diagonal resistivity, and $R_s$ is the anomalous Hall coefficient. The OHE contributions of metallic materials with high electron carrier densities are negligible compared with their AHE contributions. Thus, $\rho_{xy}$ is dominated by the AHE term and is often simply called the "anomalous Hall resistivity."

The mechanisms underlying the AHE can be classified into two categories. One is an extrinsic mechanism caused due to the scattering of conduction electrons by impurities with spin-orbit interactions, i.e., skew-scattering[4,5] and side-jump[6] mechanisms. To enhance the AHE via an extrinsic mechanism, alloying, doping, and multilayering of heavy elements with large spin-orbit interactions, such as Pt, have been demonstrated.[7–10] The other mechanism is an intrinsic mechanism due to the electronic band structure of the material.[11,12] The intrinsic AHE is related to the Berry curvature in the momentum space originating from the electronic structure of the material. Analogous to classical electromagnetism in real space, the Berry curvature is often regarded as virtual magnetic field for conduction electrons in the momentum space.[13]

Historically, extrinsic mechanisms have been considered to play a major role in the AHE. In the 2000s, however, advances in the theoretical study of the AHE by first-principles calculations helped explain the AHE in various magnetic materials through the intrinsic mechanism.[14] This led to the prediction of the AHE based on the electronic structure and in the design of materials with giant AHE. A notable example of a material with a large AHE owing to the intrinsic mechanism is the $Co_2MnAl$ ferromagnetic Heusler alloy. Chen et al. (2004) reported that a $Co_2MnAl$ epitaxial thin film exhibited $\rho_{xy} = 5$ $\mu\Omega$ cm at 300 K,[15] which was significantly greater than those of 3$d$ ferromagnetic transition metals (0.18 $\mu\Omega$ cm for Fe at 320 K[16], and 0.15 $\mu\Omega$ cm for Ni at 290 K[17]). Vilanova Vidal et al. (2011) reported an even larger $\rho_{xy}$ (>20 $\mu\Omega$ cm) in B2-ordered single-crystalline $Co_2MnAl$ thin films.[18] They showed very small temperature dependences of $\rho_{xy}$ and sensitivity $d\rho_{xy}/dH$ at temperatures between 50 °C and 300 °C, which is important for magnetic sensor applications in high-temperature environments, such as in automotive applications. Several theoretical studies have reported that the giant AHE in $Co_2MnAl$ is explained by the intrinsic mechanism.[19–21] The AHE was investigated by means of first-principles calculations, the Weyl points[20] or gapped nodal lines[21] in the band structure were found to yield a large Berry curvature, which is believed to result in a giant AHE. Materials with Weyl points are often called *Weyl semimetals*, whereas materials with the gapped nodal lines do not share a common term. In this paper, we refer to both types of materials with giant AHE as *topological magnetic materials*. Furthermore, although AHE was widely known for ferromagnets and ferrimagnets with large spontaneous magnetization, Chen et al. (2014)[22], and Kübler and Felser (2014)[23] theoretically predicted large AHE also in noncollinear antiferromagnets, such as $Mn_3Ir$, $Mn_3Ge$, and $Mn_3Sn$. As described in Ref.[24], the magnetic structure of the noncollinear antiferromagnets is not symmetric with respect to the time-reversal and translational operations. Therefore, the integrated value of the Berry curvature over the Brillouin zone, namely the anomalous Hall conductivity, can be large. This is in sharp contrast to the collinear antiferromagnets, where the existence of such symmetry leads to a complete cancellation of the Berry curvature in the Brillouin zone. Nakatsuji et al. (2015) observed a large AHE in $Mn_3Sn$ of $\rho_{xy}$ = 4 $\mu\Omega$ cm at 300 K despite its extremely low spontaneous magnetization of $\sim 7\times 10^{-3}$ $\mu_B$/f.u.,[25] which is explained to be attributed to the Weyl points in its band structure.[23] These works led to the discovery of materials with a giant AHE based on the intrinsic mechanism. To date, the giant intrinsic AHE has been reported also in a ferromagnetic $Co_2MnGa$ Heusler alloy,[26,27] ferromagnetic $Fe_3Sn_2$ and $Co_3Sn_2S_2$ kagome-compounds,[28–31], and ferromagnetic $Fe_3Ga$ and $Fe_3Al$ with the $D0_3$-order structures.[32] The largest AHE at room temperature has been reported for an $L2_1$-ordered $Co_2MnAl$ bulk single crystal with $\rho_{xy}$ = 36.9 $\mu\Omega$ cm.[21] Table 1 summarizes the physical properties of various materials including their AHE.

The application of the AHE for magnetic field sensing has attracted significant interest.[8,33–39] In a Hall cross geometry fabricated with a ferromagnetic thin film with an in-plane easy axis, the out-of-plane demagnetizing field determines $H_s$, sensitivity ($S = V_H/H_s$), and dynamic range ($DR = 2H_s$) of the AHE sensor. Therefore, in principle, $S$ and $DR$ are in conflict. Highly sensitive AHE sensors have been developed by controlling the perpendicular magnetic anisotropies of FePt and CoFeB comparable to the out-of-plane demagnetizing field, which effectively reduces $H_s$.[33–35,38] Peng et al. reported a CoFeB-based AHE sensor with a very high $S$ of 28,282 $\Omega$/T and a $DR$ of ~3 mT.[35] This $S$ value is higher than that ($S$ = 20,000 $\Omega$/T; $DR$ > 50 mT) of the commercial InSb-based OHE sensor produced by Asahi Kasei Microdevices Corporation.

The advantages of AHE sensors over semiconductor-based OHE sensors are the lower sensor resistance, lower noise level,[38,39] and smaller temperature dependence, all of which are related to the higher electron carrier density in metallic materials than in semiconductors. The lower resistance and noise characteristics of AHE sensors are particularly important for magnetic sensors with small dimensions, where the signal-to-noise ratio (SNR) of the sensor is dominated by $1/f$ noise at low frequencies and Johnson noise at high frequencies. Because of their low $1/f$ noise characteristics, AHE sensors have been proposed for the detection of weak magnetic fields with low frequencies generated by small objects such as magnetic nanoparticles and cells with magnetic markers.[40]

Another notable example of a magnetic sensor with small dimensions is the read head sensor (reader) used in hard disk drives (HDDs). Readers are magnetic sensors with dimensions of a few tens of nm that can playback recording patterns on magnetic media, for which tunnel magnetoresistance (TMR) devices have been used since the mid-2000s.[41–45] Considering that the AHE is suitable for magnetic sensors with small dimensions, AHE sensors can be potentially used as HDD readers. Figure 1(c) shows the working principle of an AHE reader. The reader flies on a rotating disk and senses high-frequency magnetic fields from the recording patterns, up to a few GHz (see the Appendix). The AHE reader comprises a cuboidal ferromagnet. A bias current is applied in the $x$ direction, and $V_H$ is measured in the $y$ direction. The magnetization of the ferromagnet is stabilized in the $x$ direction and rotated about the $z$ direction under the application of magnetic field from perpendicularly magnetized recording bits. Such an AHE reader concept was patented in 2004.[46] However, there have been no reports on the study of AHE reader. This is because the output voltages of AHE sensor devices using conventional ferromagnetic materials, such as Fe, Co, Ni, and their alloys, are so low that the obtained signal-to-noise ratio is insufficient for HDD applications. Hence, the development of materials with greater AHE is crucial for realizing AHE readers.

To date, the giant AHE in topological magnetic materials has attracted wide research interest mainly in the field of solid-state physics. The giant AHE also opens up new practical applications that are otherwise not possible with other magnetic sensor principles or with conventional materials exhibiting the AHE. In this article, we focus on the giant AHE in topological magnetic materials for applications into future reader technologies. We investigated the characteristics of AHE readers using micromagnetic and finite element method simulations and SNR estimation. The AHE readers showed promising results compared with the current TMR reader technology. Further, we highlight a fundamental challenge in the development of AHE readers: the reduction in the output voltage due to the four-terminal AHE reader geometry, and its potential solutions are also proposed. Finally, we discuss future research directions for the realization of AHE readers.

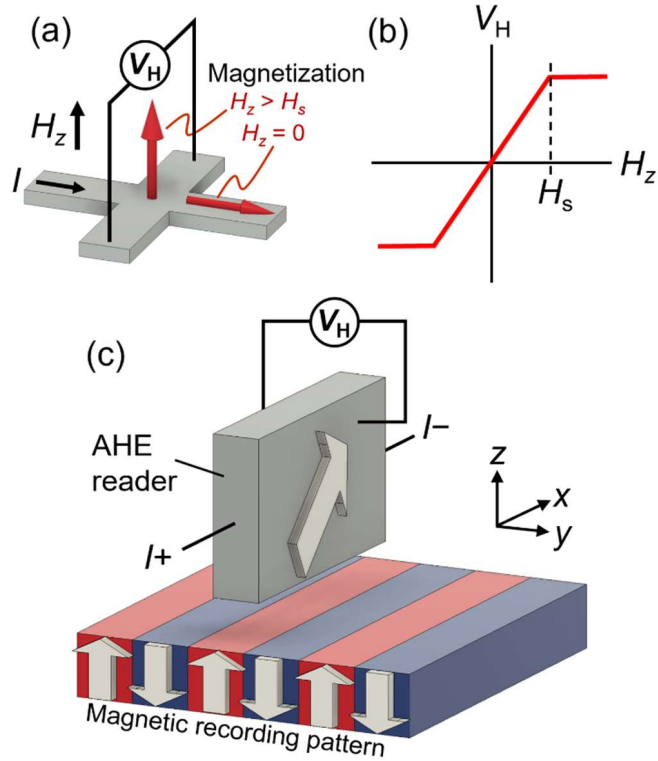

**FIG. 1.** (a) Hall cross geometry used to measure the Hall voltage ($V_H$) in the transverse direction to the bias current ($I$) and the out-of-plane magnetic field ($H_z$), and (b) corresponding $V_H$–$H_z$ curve of a ferromagnet exhibiting the AHE. (c) Schematic of the AHE reader geometry used to read out the magnetization direction of the magnetic recording pattern of an HDD.

**Table 1.** AHE and other physical properties of 3$d$ transition metals (Fe and Ni) and materials exhibiting large AHE. $T_C$: Curie temperature, $T_N$: Néel temperature, $M_s$: saturation magnetization, $\rho_{xx}$: longitudinal resistivity, $\rho_{xy}$: anomalous Hall resistivity, $\theta_H$: Hall angle ($\rho_{xy}/\rho_{xx}$), SC: single-crystal, NC: nanocrystal, and $t$: film thickness. The measurement temperature for $M_s$, $\rho_{xx}$, and $\rho_{xy}$ was 300 K unless otherwise specified. For the bulk single-crystalline Co$_2$MnAl, the data with asterisk (*) and dagger (†) were measured by applying a current parallel to the (001) and (111) planes, respectively.

| Material | Sample form | Structure | Magne-tism | $T_C$, $T_N$ (K) | $\mu_0 M_s$ (T) | $\rho_{xx}$ (μΩ cm) | $\rho_{xy}$ (μΩ cm) | $\theta_H$ | Ref. |
|---|---|---|---|---|---|---|---|---|---|
| Fe | SC film, $t$ = 93 nm | BCC | FM | 1043 | 2.15 (290 K) | 14 (320 K) | 0.18 (320 K) | 0.013 | 16 |
| Ni | Film, $t$ = 100 nm | FCC | FM | 627 | 0.61 (290 K) | 21 (290 K) | 0.15 (290 K) | 0.007 | 17 |
| Co$_2$MnAl | SC film, $t$ = 70 nm | B2 | FM | n/a | 1.08 | n/a | 22.5 (5 K) | | 18 |
| | Bulk SC | L2$_1$ | FM | 726 | 1.00 (4.2 K) | 117*, 174† | 17.9*, 36.9† | 0.15*, 0.21† | 21,47 |
| Co$_2$MnGa | Bulk SC | L2$_1$ | FM | 694 | 0.97 (5 K) 0.92 (300 K) | 120 | 15 | 0.125 | 27,48 |
| | SC film, $t$ = 50 nm | L2$_1$ | FM | n/a | 0.99 (10 K) 0.92 (300 K) | 220 | 20 | 0.09 | 49 |
| Fe$_3$Sn$_2$ | Bulk SC | kagome | FM | 670 | 0.81 (2 K) 0.73 (300 K) | 200 | 6.5 | 0.03 | 30 |
| Fe$_{0.6}$Sn$_{0.4}$ | NC film, $t$ = 40 nm | kagome+ amorphous | FM | n/a | 0.75 | 218 | 9.1 | 0.042 | 37 |
| Ta-doped FnSn | NC film, $t$ = 50 nm | kagome+ amorphous | FM | n/a | 0.31 | 369 | 11.4 | 0.031 | 50 |
| Fe$_{0.74}$Sn$_{0.26}$ | Film, $t$ = 46 nm | amorphous | FM | n/a | 1.25 | 213 | 11.9 | 0.056 | 51 |
| Mn$_3$Sn | Bulk SC | kagome | Noncollinear AFM | 420 | | 300 | 4 | 0.013 | 25,52 |
| Fe$_3$Ga | Bulk SC | D0$_3$ | FM | 720 | 1.48 | 81 | 3.5 | 0.043 | 32 |
| Fe$_{0.84}$Ga$_{0.16}$ | SC film, $t$ = 40 nm | BCC | FM | n/a | 1.68 | 119 | 5.9 | 0.050 | 53 |
| Fe$_3$Al | Bulk SC | D0$_3$ | FM | 600 | 1.39 | 93 | 4.3 | 0.046 | 32 |
| Fe$_{0.81}$Al$_{0.19}$ | SC film, $t$ = 35 nm | BCC | FM | n/a | 1.43 | 111 | 6.6 | 0.059 | 54 |
| GdFeCo | Film, $t$ = 70 nm | Amorphous | ferri | | | | 9.1 | | 55 |

## II. Magnetic recording and requirements for reader technology

The amount of digital data generated worldwide is increasing every year, with over 100 zeta bytes (= $10^{23}$ B) of data expected to be generated in 2023. Under this explosion of data, HDDs will continue to be the primary data storage device. Approximately 80% of the total storage capacity produced is by HDDs, and the rest is by solid-state drives and magnetic tapes.[56,57] Therefore, improving the areal density (AD) of the magnetic recording in HDDs is crucial for the development of a sustainable society in the era of big data. Energy-assisted magnetic recording technologies, such as heat-assisted magnetic recording (HAMR)[58–60] and microwave-assisted magnetic recording (MAMR)[61–64], have demonstrated an ultrahigh AD up to over 2 Tbit/in$^2$, which is approximately twice of that of today's HDDs. This requires the reader technology to have superior reading resolution and SNR.[65,66] Therefore, new reader technologies must be developed.

Figure 2(a) shows a schematic of the TMR readers used in today's HDDs.[67,68] A microfabricated spin-valve structure[69] with a CoFeB/MgO/CoFeB magnetic tunnel junction (MTJ)[43–45] is surrounded by top, bottom, and side magnetic shields made of NiFe soft-magnetic films. Magnetic shields are fabricated to absorb the magnetic flux from adjacent recording bits. While the magnetizations of the pinned layer and reference layer (RL) are fixed in the $z$ direction, that of the free layer (FL) is stabilized in the $x$ direction by the stray magnetic field from the side shields. When a magnetic field from the recording media (media field) is applied in the $z$ direction, the FL magnetization is rotated about the $z$ direction, and the output voltage is approximately proportional to the strength of the media field. Figure 3 shows a calculated output voltage ($V$)−media field ($H_z$) transfer curve of a TMR reader with a TMR ratio of 100%. TMR ratio is defined as $(R_\text{AP} - R_\text{P})/R_\text{P}$, where $R_\text{P}$ and $R_\text{AP}$ are the resistances in the parallel and antiparallel magnetization states between the FL and RL. To ensure a fast and stable response of the FL magnetization to the high-frequency media field, TMR readers operate in the range of the pseudo-linear $V$–$H_z$ curve, as shown by the shadowed region in Fig. 3. Then, reader utilization, $\eta$, is defined by

$$\eta = \frac{V_{\max} - V_{\min}}{V_{\text{AP}} - V_{\text{P}}}, \qquad (1)$$

where $V_{\max}$ and $V_{\min}$ are the maximum and minimum voltages under the positive and negative media fields, respectively, and $V_{\text{AP}}$ and $V_{\text{P}}$ are the voltages in the parallel and antiparallel magnetization states, respectively.[67] The typical value of $\eta$ for the TMR readers is 0.3, for which the FL magnetization is rotated approximately by ±20° about the magnetic easy axis in the $x$ direction.

The important parameters of a TMR reader are the TMR ratio, resistance-area product (*RA*), and dimensions of the TMR sensor in terms of the output voltage, noise, and spatial resolution of the reader, respectively. Compared with the MTJs using amorphous $AlO_x$ and $TiO_x$ tunnel barriers that were implemented in HDDs in mid-2000s,[41,42] the CoFeB/MgO/CoFeB MTJs can achieve an ultralow *RA* value of ~0.25 $\Omega\,\mu m^2$ with a high TMR ratio of 120%.[70] The physical spacing between the side shields (reader width (*W*)) and that between the top and bottom shields, the so-called read gap (*G*) or shield-to-shield spacing, are the major factors determining the spatial resolution of the reading. Thus, *W* and *G* should be reduced to achieve higher ADs. *W* is determined by the nanofabrication process using photolithography and Ar ion milling, and *G* is determined by the total thickness of the reader film. The values of *W* and *G* of the existing TMR readers for an AD of 1 Tbit/in$^2$ are ~25 nm. Table 2 shows the predictions of *W* and *G* for an AD of 2.4 Tbit/in$^2$ and higher; clearly *W* should be below 12 nm and *G* should be below 20 nm for the next-generation HDDs.[66]

TMR readers are associated with two fundamental problems when it comes to reducing the reader dimensions *W* and *G*. First, the lowest *RA* value ever reported of 0.25 $\Omega\,\mu m^2$ is still too high for $W < 20$ nm in terms of the reader resistance (impedance) and thermal noise (Johnson noise and/or shot noise); the optimal $RA$ value for an AD of 2.4 Tbit/in$^2$ is predicted to be 0.1 $\Omega$ $\mu$m$^2$.[66] In addition, the reader impedance ($R$) matters the impedance matching with the amplifier and the signal transmission from the reader to the amplifier and signal processor. The latter is because $R$ and the stray capacitance of the transmission line form an $RC$ circuit (low-pass filter), which attenuates the high-frequency readback signal.[65,71] Reducing the $RA$ of the TMR reader by thinning the MgO tunnel barrier often results in a rapid decrease in the TMR ratio and a rapid increase in the ferromagnetic interlayer coupling between the CoFeB RL and FL through the MgO tunnel barrier, which hinders the reader operation. Second, it is difficult to fit a multilayered spin-valve structure for $G$ below 20 nm. To overcome these challenges, alternative sensor devices have been proposed, *e.g.*, all-metallic current-perpendicular-to-plane giant magnetoresistance (CPP-GMR) sensors[72–75] with $RA$ ~0.05 $\Omega$ $\mu$m$^2$ and nonlocal spin-valve sensors[76–79] that eliminate antiferromagnetic and pinned layers from the read gap. However, none of these methods have simultaneously demonstrated sufficient resolution and SNR.

Figure 2(b) shows a schematic of the AHE reader, which is mainly composed of a single ferromagnetic film as the sensing layer (SL). Nonmagnetic contact layers are inserted between the SL and the shields to decouple the direct exchange coupling between them. The magnetization of the SL is stabilized (biased) by the stray magnetic field from the side shields in the $x$ direction. The SL magnetization is rotated in the $z$-$x$ plane when the magnetic field from the recording media is applied in the $z$ direction. The $RA$ value of the AHE readers is expected to be less than 0.05 $\Omega$ $\mu$m$^2$ when a 20 nm-thick film with $\rho_{xx} = 200$ $\mu\Omega$ cm is used for the SL. Such a small $RA$ value leads to reduction in the Johnson noise and the amplifier noise of the reader (See the Appendix). In addition, because of the simple structure of the AHE reader, thick

SLs can be used within the permitted space of *G* compared with the FLs of the TMR reader. Thicker SLs are expected to suppress noise by thermal magnetization fluctuation (mag-noise), which is dominant in readers with small dimensions.[65,66] Table 3 presents comparisons of the AHE and other sensor technologies for general magnetic sensor and reader applications. The simple structure of the AHE sensor is advantageous for HDD reader applications. However, realizing a sufficiently high output voltage in the reader structure is a major challenge, as discussed in detail in the following sections.

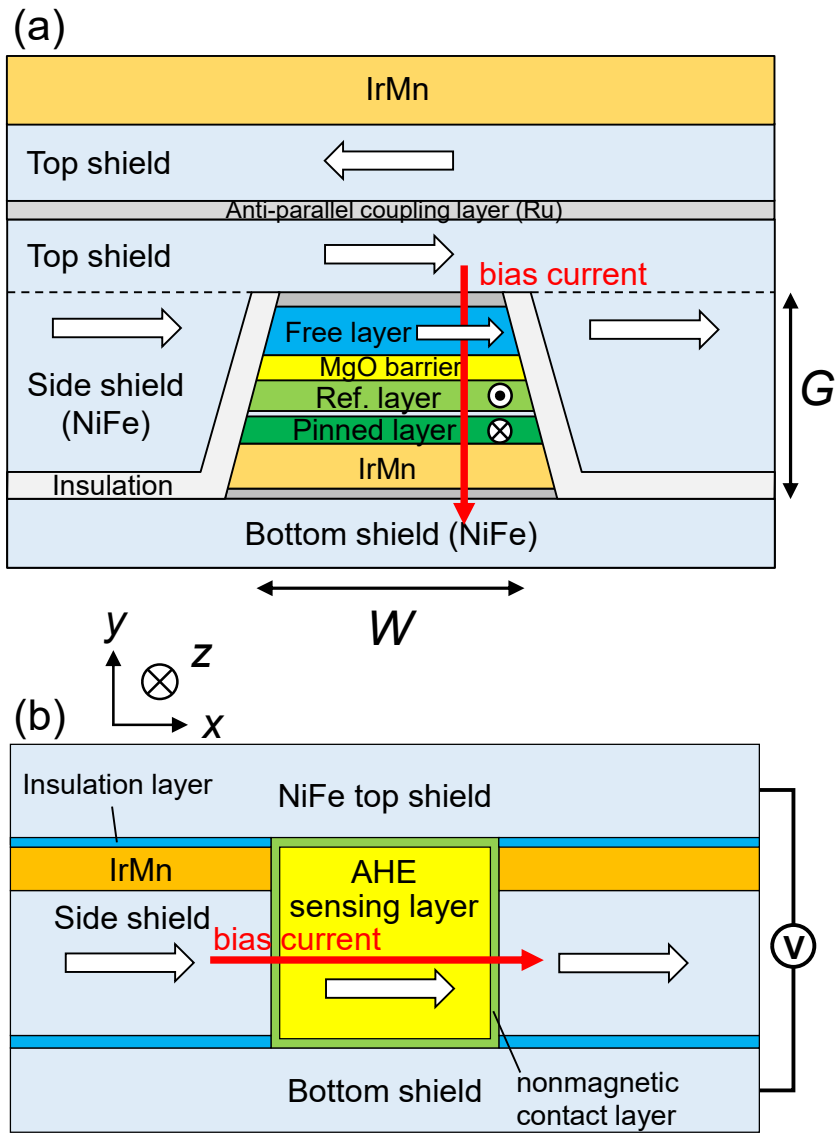

**FIG. 2.** Schematic structures of (a) existing TMR reader and (c) proposed AHE reader seen

from the magnetic recording media side (air bearing surface). The white arrows in the shields and free (sensing) layer indicate their magnetization directions.

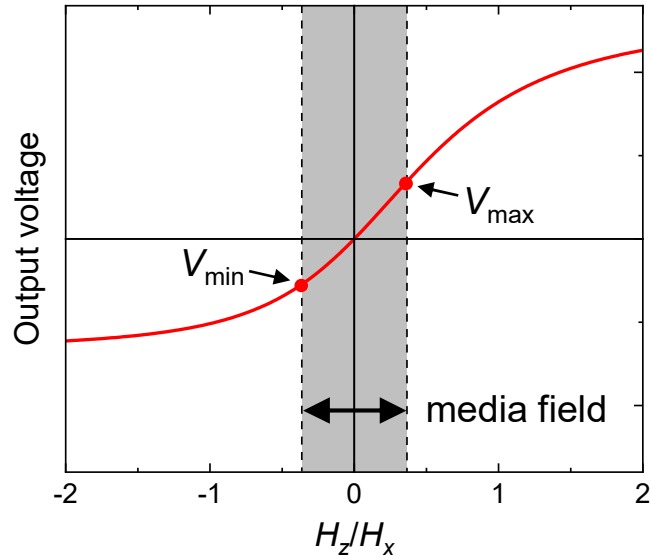

**FIG. 3.** Calculated output voltage−media field ($H_z$) transfer curve of a TMR reader with a TMR ratio of 100%. $H_z$ is normalized by the strength of the bias magnetic field ($H_x$). The shadowed region corresponds to the reader utilization, $\eta = 0.3$.

**Table 2.** Predicted physical dimensions (reader width and gap) of the reader for Ads in the range of 2.4–6.0 Tbit/in$^2$.[66] The thicknesses of TMR FL and AHE SL used for the simulations and SNR estimations.

| Areal density (Tbit/in$^2$) | 2.4 | 4.0 | 6.0 |
|---|---|---|---|
| Reader width, $W$ (nm) | 12 | 9.0 | 6.0 |
| Reader gap, $G$ (nm) | 20 | 17 | 14 |
| Thickness of AHE SL (nm) | 16 | 13 | 10 |

| Thickness of TMR FL (nm) | 5 | 4 | 3 |

Table 3 Comparison of the various magnetic sensor technologies. CIP: current-in-plane, CPP: current-perpendicular-to-plane, GMR: giant magnetoresistance.

|  | **OHE** | **AHE** | **CIP-GMR** | **TMR** | **CPP-GMR** | **Nonlocal spin-valve** |
|---|---|---|---|---|---|---|
| **Terminal geometry** | Four-terminal Hall cross | | Two-terminal CIP | Two-terminal CPP | | Four-terminal, spin injection and detection separated |
| **Key materials** | Narrow gap semiconductors (e.g., InSb) | Topological magnetic materials (e.g., $Co_2MnAl$, $Co_2MnGa$) | fcc-CoFe/Cu/CoFe | CoFeB/MgO/CoFeB | Half-metallic Heusler alloys | |
| **Advantages for general magnetic sensors** | High sensitivity, wide dynamic range, and low cost | Low noise, and small temperature dependence | Moderately large MR ratio (~20%), and low cost | Large MR ratio (~200%), wide range of resistance, and wide controllability of dynamic range | n/a | n/a |
| **Challenges for general magnetic sensors** | High noise, large temperature dependence | Sensitivity and dynamic range in conflict | Narrower controllability of dynamic range than TMR | Higher manufacturing cost than CIP-GMR | Small sensor resistance and output voltage | Small output voltage |
| **Advantages for HDD readers** | n/a | Compatible to small dimensions, and high SNR potential | Larger output voltage than anisotropic magnetoresistance | Large output voltage | Low RA and low noise | Compatible to small dimensions |
| **Challenges for HDD readers** | High noise | Output voltage reduction in reader geometry | CIP geometry is unsuitable for small dimensions. | Difficult to reduce RA to 0.1 Ω μm$^2$, reducing total thickness, and mag-noise dominant | Smaller output than TMR, reducing total thickness, and mag-noise dominant | Small output voltage, and difficult fabrication process |

## III. Methods

Theoretical and simulation methods were employed in our investigation on the AHE reader. The SNRs of the TMR and AHE readers were calculated analytically from the theoretical expressions for the output voltage and noise. Different types of noises were considered: Johnson noise, shot noise, thermal magnetic noise, and amplifier noise. The equations and physical parameters for the calculations of the output voltage and noises are described in the Appendix. The magnetization properties of the TMR FL and AHE SL were simulated by micromagnetic simulations using the COMSOL Multiphysics® with a micromagnetic module,[80] and also by the Object Oriented MicroMagnetic Framework (OOMMF).[81] The electric current distribution and Hall voltage in the AHE readers were studied using the finite element method (FEM) simulations in COMSOL Multiphysics.

## IV. Magnetization and output voltage response curves

### A. Magnetization curves

The output voltage vs. the magnetic field transfer curve is one of the most fundamental properties for HDD reader. In this section, we describe the responses of the AHE and TMR readers to a magnetic field using micromagnetic simulations. We only considered the magnetizations of the SL and FL of the AHE and TMR readers, respectively, whereas the magnetizations of the pinned and reference layers of the TMR reader (Fig. 2(a)) were considered to be fixed within the magnitude of the magnetic field. The effect of the magnetic shields was considered only as a bias magnetic field in the $x$ direction. For both the SL and FL, we assumed common magnetic properties: saturation magnetization ($M_s$) of 1000 emu/cm$^3$ (≈1.25 T) and zero magneto-crystalline anisotropy. Table 2 presents the reader width ($W$) and gap ($G$) for ADs of 2.4–6.0 Tbit/in$^2$ predicted by Albuquerque et al.[66]. For simplicity, the FL and SL dimensions in the z direction, that is, the stripe height (*SH*), were assumed to be identical to *W*. Thus, SL and FL do not exhibit shape anisotropy in the *z-x* plane. The primary difference between the AHE and TMR readers is the thickness of the ferromagnetic SL (FL). Because of the simpler structure of the AHE reader, the SL can be significantly thicker than the FL of the TMR reader. As shown in Table 2, we assumed SL thicknesses to be 16, 13, and 10 nm for ADs of 2.4, 4.0, and 6.0 Tbit/in$^2$, respectively, corresponding to (*G*−4) nm considering that the thickness of both the seed and cap layers is 2 nm. On the other hand, the FL thicknesses were assumed to be 5, 4, and 3 nm for ADs of 2.4, 4.0, and 6.0 Tbit/in$^2$, respectively. For micromagnetic simulations, the mesh size of the ferromagnetic layer was maintained below 2 nm.

Figures 4(a) and (c) show the schematics of the SL and FL of the AHE and TMR readers, respectively, with the dimension of AD = 4.0 Tbit/in$^2$. The bias magnetic field ($H_x$) from the side shield was applied in the +*x* direction, and the magnetic field (stray field from the magnetic media) was applied in the ±*z* direction. Figure 4(b) shows the magnetization components vs. $H_z$ of the SL of the AHE reader. Because of the shape magnetic anisotropy in the SL ($t_{SL} > W = SH$), the magnetic easy axis of the SL magnetization was in the *y* direction with an anisotropy field ($H_k$) of ~140 mT, by which the $m_z$–$H_z$ curve was linear without a bias magnetic field ($H_x$ = 0). Figure 4(b) shows the case with a bias field of $\mu_0 H_x$ = 50 mT, which shows only a slight change in the $m_z$–$H_z$ curve compared with that for $H_x$ = 0. Thus, the AHE reader with the dimensions of $t_{SL} > W$ functions without a bias field, which can be an advantage for the freedom of the reader structure design. The value of $H_k$, can be controlled by $t_{SL}$: $\mu_0 H_k$ = 40, 75, and 110 mT for $t_{SL}$ = 10, 11, and 12 nm.

Figure 4(d) shows the magnetization curves of the FL of the TMR reader. Because the dimensions were $t_{FL}$ = 4 nm and *W* = *SH* = 9 nm, the FL was magnetized in the film plane (*z-x* plane). For $H_x = 0$, the $m_z$–$H_z$ curve showed a magnetization reversal at low $H_z$ values. Thus, it cannot be used as an HDD reader. When a bias field of $\mu_0 H_x \sim 50$ mT was applied, $m_z$ gradually rotated with increasing $H_z$, which was suitable for magnetic field sensing. Therefore, applying a lateral bias field is essential for in-plane magnetized FLs that are used by the commercial readers.

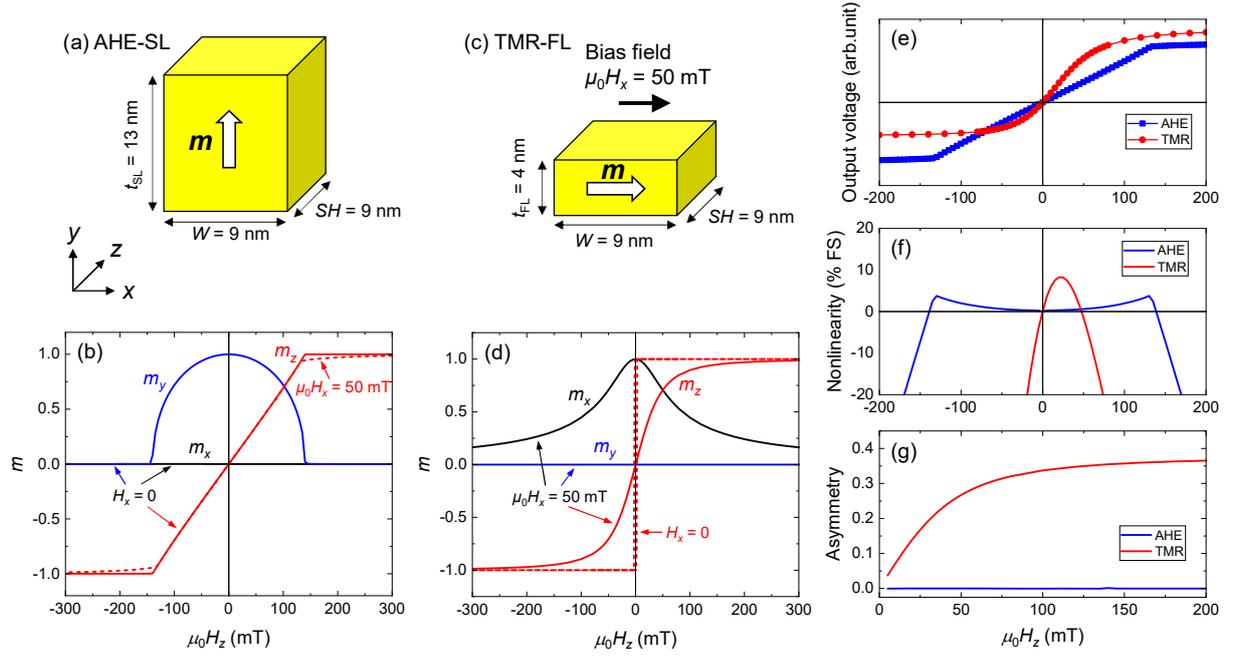

**FIG. 4.** (a) and (c) Schematics of the SL and FL and (b) and (d) Magnetization curves of the AHE and TMR sensors, respectively. (e) Output voltage ($V$)–$H_z$ transfer curves, (f) nonlinearity, and (g) asymmetry of the AHE and TMR readers.

## B. Output voltage transfer curves

The $m_z$–$H_z$ curves were converted to output voltage ($V$)–$H_z$ curves, as shown in Fig. 3(e). For the AHE readers, the Hall voltage measured in the $y$ direction is proportional to $H_z$. Thus, the conversion of the $m_z$–$H_z$ curve to the $V$–$H_z$ curve is straightforward, and both have identical curve shapes with linear dependences of $m_z$ and $V$ on $H_z$. For the TMR readers, the conversion of the $m_z$–$H_z$ curve into the $V$–$H_z$ curve is more complicated. The conductance ($G$) of the TMR sensor is calculated as follows.

$$G = \frac{G_{\max}+G_{\min}}{2} + \frac{G_{\max}-G_{\min}}{2} \cdot \cos\theta_{\mathrm{FL-RL}}, \tag{2}$$

where $G_{\max}$ and $G_{\min}$ are the maximum and minimum values of $G$ in the P and AP magnetization states, respectively, and are related to the TMR ratio as TMR ratio $= \frac{G_{\min}-G_{\max}}{G_{\max}}$. $\theta_{\mathrm{FL-RL}}$ is the relative angle of the magnetizations of FL and RL, which is calculated by the micromagnetic simulations. With this, $V$ can be calculated as follows:

$$V = \Delta R \cdot I_b = (1/G - 1/G_{H_z=0}) \cdot I_b, \tag{3}$$

where $I_b$ is the constant bias current flowing through the TMR reader. In the present case, we assumed the TMR ratio to be 120%, which is typical for current ultralow $RA$ TMR devices.[70] As shown in Fig. 3(e), the $V$–$H_z$ curve of the TMR reader is asymmetric with respect to $H_z = 0$, which is generally obtained for TMR devices because of the relationship represented by Eqs. (2) and (3).

Figures 3(f) and (g) show the nonlinearity and asymmetry of the $V$–$H_z$ curves for both types of readers, respectively. The nonlinearity is defined by $[V(H_z) - V_{\mathrm{fit}}(0)]/V(H_z)$, where $V_{\mathrm{fit}}(0)$ is the fitted line for the $V(H_z)$ curve at $H_z = 0$, and the asymmetry is defined by $\frac{V(H_z)+V(-H_z)-2V(0)}{V(H_z)-V(-H_z)}$.[67] In general, nonlinearity and asymmetry are important characteristics of magnetic field sensors because they affect the accuracy of the magnitude of the magnetic field, which is also the case for HDD readers. While the $V$–$H_z$ transfer curve of the AHE reader shows a nonlinearity of up to 4% full-scale (FS) for $H_z < H_s$, that of the TMR reader shows a significantly greater nonlinearity. Similarly, the asymmetry of the transfer curve of the AHE reader was zero, whereas that of the TMR reader was significantly greater. Because of the nonlinear and asymmetric $V$–$H_z$ transfer curves of the TMR readers, the reader utilization, $\eta$, is typically limited to 0.3. Limiting $\eta$ to 0.3 also ensures a stable magnetization response of the FL to the GHz-range AC magnetic field from the recording media. However, the perfect symmetry and low nonlinearity of the AHE readers may allow the application of a larger $\eta$ than the conventional value for TMR readers, which produces a larger signal and an improved reader SNR. This is a significant advantage of the AHE reader over the TMR reader.

Note that a DC offset voltage caused by $\rho_{xx}$ and a voltage due to the ordinary magnetoresistance (OMR) of the SL material, which is an even function with respect to $H_z$, are added to the $V$–$H_z$ transfer curve of the AHE reader when the configuration of the top and bottom contacts for detecting the output voltage are not completely symmetric. The anisotropic magnetoresistance of the SL material does not add voltage because the angle between the directions of the SL magnetization and the bias current is constant during the reader operation. While the DC offset voltage can be corrected by the signal processing, the voltage by OMR can cause a distortion of the $V$–$H_z$ transfer curve. However, the sensitivity of OMR in ferromagnets is generally small, e.g., ~1.4 %/T for the $Co_2MnAl$,[21] and $H_z$ is only up to several tens of mT. Therefore, the distortion of the $V$–$H_z$ transfer curve by OMR may be negligibly small.

**V. Signal-to-noise ratio estimation**

We estimated the SNR of the AHE and TMR readers by analytically calculating the signal and noise. In AHE readers, assuming that the bias current flows only in an SL with thickness $t_{SL}$, the Hall voltage ($V_H$) at the magnetization saturation in $z$ direction is given by

$$V_H = \rho_{xy} J_x t_{SL}, \qquad (4)$$

where $J_x$ denotes the bias current density. Thus, the peak-to-peak output voltage ($V_{out}$) for a reader utilization $\eta$ is given by

$$V_{out} = 2\rho_{xy} J_x t_{SL} \eta. \qquad (5)$$

While the anomalous Hall resistivity ($\rho_{xy}$) determines $V_{out}$ as expressed in Eq. (5), the longitudinal resistivity ($\rho_{xx}$) determines the reader resistance in the $z$ direction by $R = \rho_{xx} t_{SL}/(W \cdot SH)$, Johnson noise, and amplifier noise. In addition to these types of noise, the thermal magnetic noise (mag-noise) was considered. We calculated the SNR for the $\rho_{xx}$ and $\rho_{xy}$ values reported for Fe ($\rho_{xx}$ = 14 $\mu\Omega$ cm and $\rho_{xy}$ = 0.18 $\mu\Omega$ cm)[16] and Co$_2$MnGa ($\rho_{xx}$ = 220 $\mu\Omega$ cm and $\rho_{xy}$ = 20 $\mu\Omega$ cm)[49] thin films (Table 1). $J_x$ applied to the AHE reader was assumed to be 2×10$^8$ A/cm$^2$. For the TMR readers, we assumed $RA$ = 0.25 $\Omega$ $\mu$m$^2$ and TMR ratio = 120%,[70] and the bias voltage ($V_b$) to the TMR reader was assumed to be 100 mV, which is typical for current TMR readers. The reader SNRs for the dimensions for AD = 2.4−6.0 Tbit/in$^2$ were calculated. Although $\eta$ for AHE readers may be higher than that for TMR readers due to the linear and asymmetric $V$–$H_z$ curves of AHE readers as discussed in the previous section, for simplicity, we set $\eta$ to 0.3 for both the types of readers, which is the conventional value. The impedance matching with the amplifier and the attenuation of the high-frequency readback signal were not considered for the SNR estimations. The other parameters used for the SNR calculation are described in the Appendix.

Figure 5(a) shows the SNR vs. AD. For both the types of readers, the SNR decreased with increasing AD because of the increasing noise in smaller dimensions for higher AD. The AHE reader with Fe exhibited negative SNRs, indicating that the output voltage was lower than the noise voltage. Thus, the AHE readers using conventional ferromagnets with small AHE have not been considered for the reader applications. However, Co$_2$MnGa significantly increased the SNR of the AHE reader, which was higher than that of the TMR reader. This demonstrates the potential of AHE sensors using topological magnetic materials, such as Co$_2$MnGa and Co$_2$MnAl, for future reader applications.

To understand why the AHE reader using Co$_2$MnGa was predicted to show a higher SNR than the TMR reader, we examined the magnitude of each noise component in both the types of readers. Figures 5(b) and (c) show the noise voltage densities of the TMR and AHE readers using $Co_2MnGa$, respectively. The noise in the TMR readers is dominated by mag-noise, owing to the small volume (thin thickness) of the FLs. The domination of mag-noise means that an increase in the TMR ratio does not improve the SNR because the amplitude of the mag-noise linearly scales with the output voltage.[82,83] Moreover, since both the shot noise ($\propto \sqrt{R}$) and amplifier noise ($\propto R$) have small contributions compared with the mag-noise, a reduction in $RA$ has a minor impact on the SNR. Thus, SNR improvement can only be achieved by reducing the mag-noise, which can be done by increasing the magnetic volume ($M_s V_{FL}$) and or decreasing the FL magnetic damping constant ($\alpha$).

In contrast, for AHE readers, the contributions of the Johnson noise, amplifier noise, and mag-noise are comparable, as shown in Fig. 5(c). Therefore, increasing $V_{out}$ by enhancing $\rho_{xy}$ can effectively improve the SNR. Figure 5(d) shows the contour plot of the SNR of the AHE reader in the 4 Tbit/in$^2$ dimension for $\rho_{xx}$ and $\rho_{xy}$, showing a higher SNR for higher $\rho_{xy}$ at a given value of $\rho_{xx}$. The $\rho_{xx}$ and $\rho_{xy}$ values reported for the $Co_2MnGa$ thin film[49] and $Co_2MnAl$ bulk sample[21] are also plotted in Fig. 5(d). The predicted SNR for the $\rho_{xx}$ and $\rho_{xy}$ values of the bulk $Co_2MnAl$ was 28.4 dB, which is significantly higher than those for the film $Co_2MnGa$ (23.9 dB) and the TMR reader (22.2 dB). Therefore, the giant AHE observed in bulk $Co_2MnAl$ is attractive for reader applications, and realizing such a high $\rho_{xy}$ in thin films is highly desired. Furthermore, considering the lower impedance of the AHE reader ($R = 353$ Ω in the dimension of AD = 4.0 Tbit/in$^2$) than the TMR reader ($R = 3086$ Ω) in terms of the impedance matching and the attenuation of high-frequency signal, we expect a further SNR advantage for the AHE reader.

As mentioned in the previous section, the $V$–$H_z$ transfer curve of the AHE reader shows significantly lower nonlinearity and essentially zero asymmetry compared with that of the TMR reader. Thus, a greater $\eta$ than the conventional $\eta = 0.3$ for TMR readers may be applied to AHE readers. Figure 5(e) shows the contour plot of the SNR for $J_x$ and $\eta$ calculated for the $\rho_{xx}$ and $\rho_{xy}$ values of the Co$_2$MnGa film. For a given $J_x$, e.g., $2\times10^8$ A/cm$^2$, an increase in $\eta$ improves the SNR; 23.9 dB and 28.4 dB for $\eta = 0.3$ and 0.5, respectively. The principle that limits the maximum $\eta$ in practical readers is expected to be the response of the SL magnetization to the ~2 GHz media field (see the Appendix). For low $\eta$, the SL magnetization responds to the media field by the magnetization rotation process about the hard axis. With increasing $\eta$, the SL magnetization response to the media field approaches the magnetization reversal process associated with the magnetization procession, which generally requires a longer time than the magnetization rotation for low $\eta$. To answer the question of how high $\eta$ is usable for AHE readers, micromagnetic simulation studies on the dynamic response of the SL magnetization should be conducted.

Figure 5(e) shows the range of $J_x$ required for AHE readers to achieve a higher SNR than that of the TMR reader. For $\eta = 0.3$ and 0.5, the SNR of the AHE reader exceeds that of the TMR reader at $J_x > 1.7\times10^8$ A/cm$^2$ and $1.0\times10^8$ A/cm$^2$, respectively. Therefore, the application of a higher $J_x$ is desirable to achieve a higher SNR in AHE readers. In metallic spintronics devices, such as CPP-GMR sensors and spin-torque oscillators, the bias current density is limited by Joule heating and the spin transfer torque (STT) effect. In CPP-GMR readers, the magnetization instability by STT, so-called the spin torque noise,[84–86] limits the bias current density to ~$1.5\times10^8$ A/cm$^2$.[87] In spin-torque oscillators, the STT is utilized to oscillate the FL magnetization in the bias current density range of $2\times10^8$–$5\times10^8$ A/cm$^2$,[62,88] where the long-term reliability of the device is dominated by Joule heating and electromigration. In the case of an AHE reader composed of a single ferromagnetic layer, the spin torque noise is considered to be negligible. Thus, high $J_x > 2\times10^8$ A/cm$^2$ may be applied to the AHE readers, and an improved

SNR can be obtained.

Another advantage of the AHE readers is that $\rho_{xy}$ of the topological magnetic materials shows a small temperature dependence, whereas the performance of TMR and CPP-GMR devices decreases with increasing temperature. The bulk $Co_2MnGa$ single crystal has been reported to have an approximately constant $\rho_{xy}$ of ~15 $\mu\Omega$ cm between 300 K and 400 K, whereas $\rho_{xx}$ increases from ~120 $\mu\Omega$ cm at 300 K to ~135 $\mu\Omega$ cm at 400 K.[27] Assuming a similar temperature dependence of $\rho_{xy}$ and $\rho_{xx}$ in the $Co_2MnGa$ thin films used for AHE readers, a temperature increase from 300 K to 400 K and the corresponding increase in $\rho_{xx}$ degrade the reader SNR by 1.3 dB.

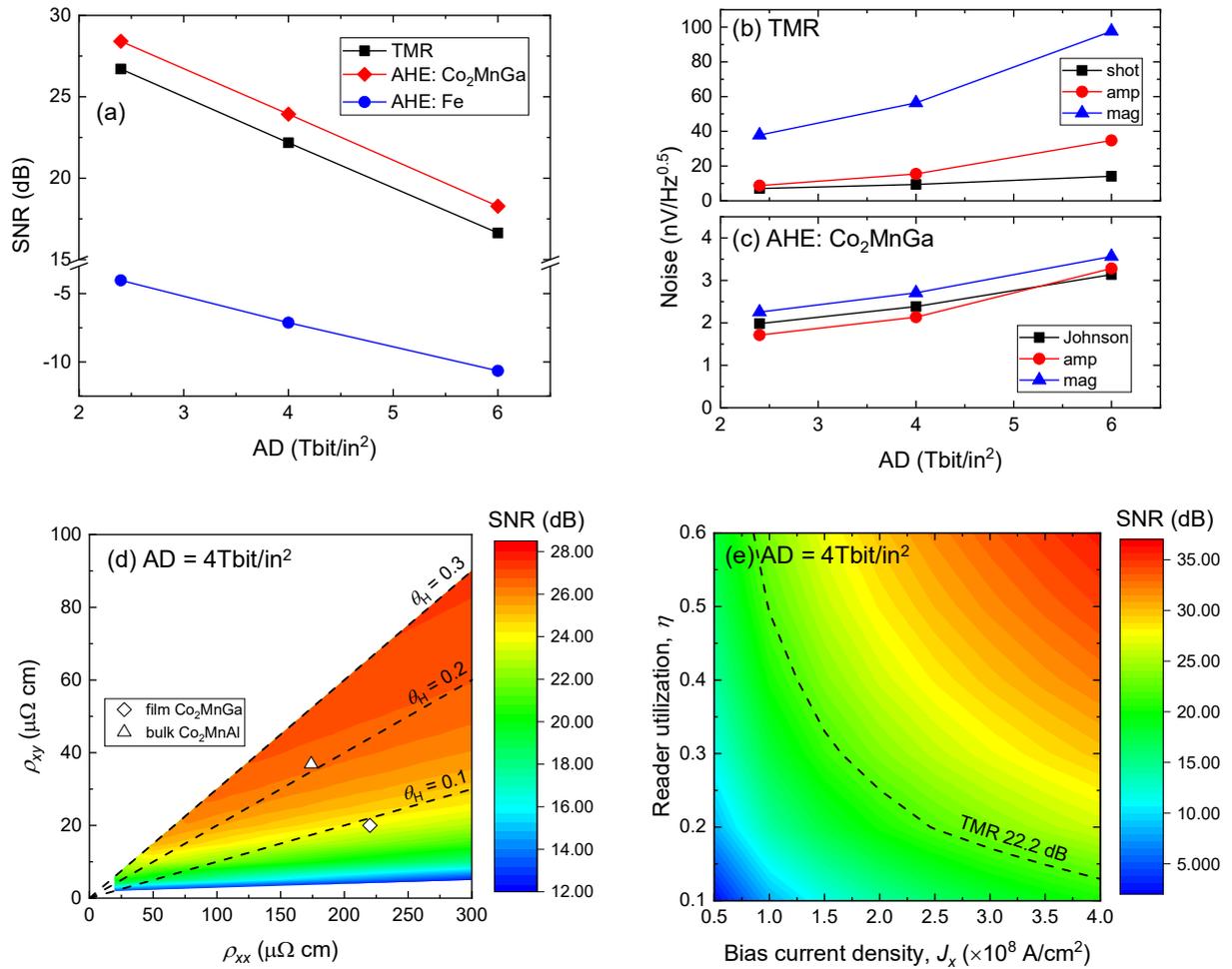

**FIG. 5.** (a) Calculated SNR vs. AD for TMR ($RA = 0.25$ $\Omega$ $\mu m^2$ and TMR ratio of 120%) and

AHE readers with $\rho_{xx}$ = 220 $\mu\Omega$ cm and $\rho_{xy}$ = 20 $\mu\Omega$ cm reported for Co$_2$MnGa.[49] (b) and (c) Voltage densities of each noise component as a function of AD for the TMR and AHE readers, respectively. Contour plots of the SNR for (d) $\rho_{xx}$ and $\rho_{xy}$, and (e) $J_x$ and $\eta$ of the AHE reader for AD = 4 Tbit/in$^2$ (Table 2). $\rho_{xx}$ = 220 $\mu\Omega$ cm and $\rho_{xy}$ = 20 $\mu\Omega$ cm are used for (e).

## VI. FEM simulations: Current distribution

### A. Output voltage reduction in reader geometry

As schematically shown in Fig. 2(b), the AHE reader is surrounded by soft magnetic shields that function as four-terminal electrodes. Thus, the bias current distribution in the AHE sensor is complicated because the shields can shunt the bias current. To study the current distribution in the AHE readers and its effect on $V_H$, we performed FEM simulations in the AHE reader structure shown in Fig. 6(a). To reduce the simulation time, the thicknesses of the bottom and top shields ($t_{NiFe}$) were kept the same as the reader width, $W$, and we confirmed that the simulation results were nearly independent of $t_{NiFe}$ when $t_{NiFe} \geq W$. We inserted 1-nm-thick nonmagnetic metallic contact layers between the SL and NiFe shields to decouple the exchange coupling between them. We assumed $SH = W$ as in the micromagnetic simulation shown in Fig. 4. The resistivity of the NiFe shields was assumed to be 20 $\mu\Omega$ cm, which is typical in experiments. For the nonmagnetic contact layers, we assumed a resistivity of 25 $\mu\Omega$ cm as Ta or Ru, which are often used for the seed or capping layers of spintronic devices. $V_H$ is defined as the voltage between the saturated magnetization states in the $x$ direction and in the $z$ direction, corresponding to the "zero-to-peak" voltage of the AHE reader for the excitation field applied in the $z$ direction. We simulated $V_H$ for the SL with $\rho_{xx}$ = 220 $\mu\Omega$ cm and $\rho_{xy}$ = 20 $\mu\Omega$ cm reported for the Co$_2$MnGa film[49] and a bias current density of $J_x = 2 \times 10^8$ A/cm$^2$ applied between the side shields.

Figure 6(b) shows $V_H$ vs. $W$ for $t_{SL}$ ranging from 5 to 35 nm. When $W$ was increased from 5 nm, $V_H$ initially increased and then decreased after reaching a peak value at $W$ = 10−35 nm. The decrease in $V_H$ with increasing $W$ can be explained by a shunting of the bias current through the top and bottom shields, as schematically shown in the inset of Fig. 6(c). Figure 6(c) shows the average value of $J_x$ in the SL for various $W$ and $t_{SL}$. Although we applied $J_x$ = $2 \times 10^8$ A/cm² between the side shields, the average $J_x$ in the SL was significantly lower and decreased with increasing $W$ due to current shunting through the top and bottom shields. Because $V_H$ is proportional to $J_x$ in the SL, the current shunting significantly decreases $V_H$. On the other hand, the decrease in $V_H$ as $W$ decreases in the range of small value of $W$ (e.g., $W \leq 35$ nm for $t_{SL}$ = 35 nm) is understood by a leakage of $V_H$ by the adjacent side shields. Figure 6(d) shows the equivalent electric circuits of the SL and side shields. Owing to the finite resistance of the metallic side shields ($R_{SS}$), $V_H$ caused a leakage current in the side shields. This decreased the observed Hall voltage $V_H^{obs}$ from the intrinsic $V_H^{int}$ as expressed by $V_H^{obs} = V_{HE}^{int} \cdot R_{SS}/(R_{SL} + R_{SS})$, where $R_{SL}$ is the resistance of the SL. When $W$ decreases, $R_{SL}$ increases while $R_{SS}$ remains constant, and thus, $V_H^{obs}$ drops as seen in Fig. 6(b) (e.g., $W \leq 35$ nm for $t_{SL}$ = 35 nm). Therefore, the shunt of the bias current and the leakage of $V_H$ in the AHE reader structure decrease $V_H$ compared with that obtained in the simple Hall cross geometry expressed in Eq. (4). For example, when $W$ = $SH$ = 10 nm and $t_{SL}$ = 15 nm, which is approximately for AD = 4 Tbit/in², $V_H$ = 3.0 mV by Eq. (4). However, the current shunting and $V_H$ leakage resulted in a significant reduction in $V_H$ to only 0.2 mV. Therefore, developing a method to suppress the reduction in output voltage of the AHE reader is indispensable.

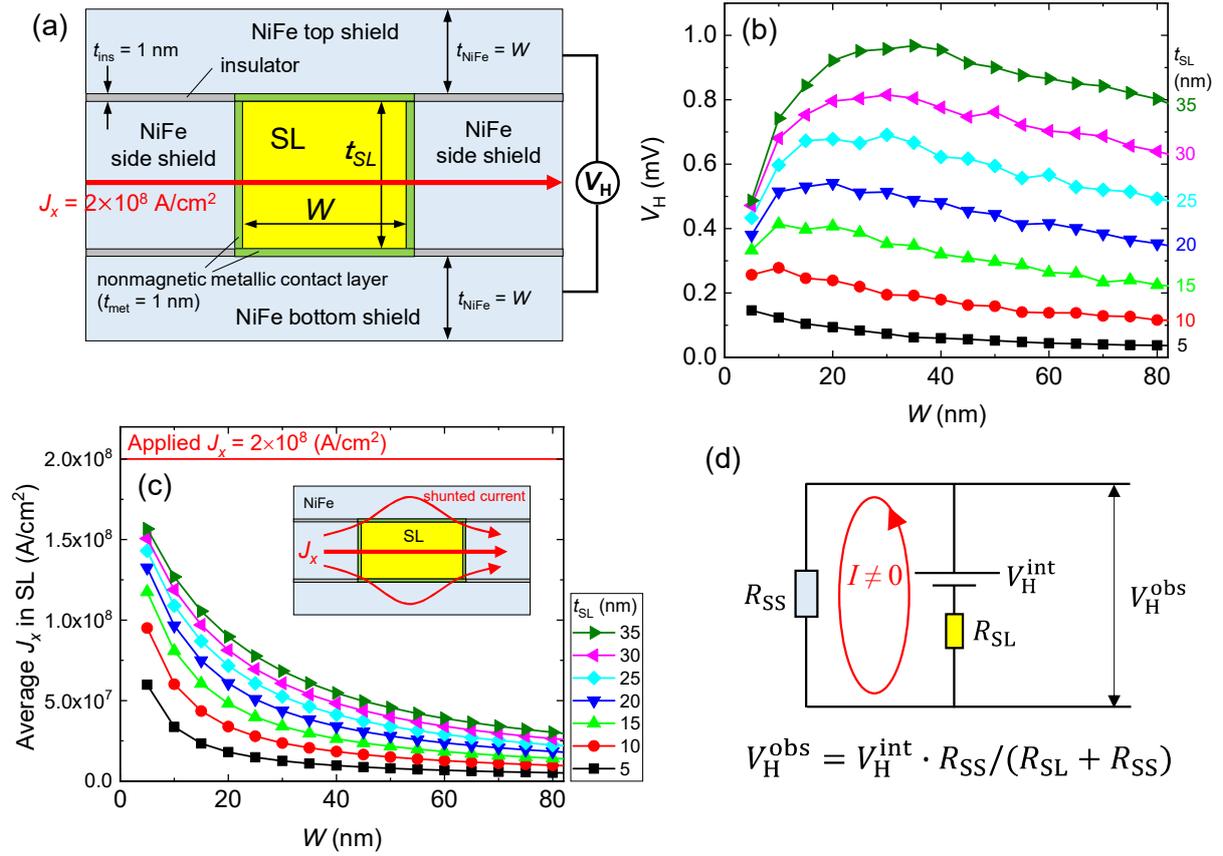

**FIG. 6.** (a) Structure of the AHE reader for FEM simulations, (b) $V_H$ vs. SL width ($W$) for various SL thicknesses ($t_{SL}$), (c) Dependence of the average bias current density ($J_x$) within the SL on $W$ when $J_x = 2 \times 10^8$ A/cm$^2$ is applied between the side shields. The inset schematically shows the bias current shunting through the top and bottom shields. (d) Equivalent electric circuit for the SL and side shields (SS), indicating a leakage of $V_H$ through the side shields.

### B. Reader structures to suppress the reduction in output voltage

To suppress the output voltage reduction by the bias current shunting and $V_H$ leakage, a simple solution is to insert highly resistive materials for the contact layers between the SL and shields. This approach suppresses the shunting and leakage based on the FEM simulations (not shown here). However, the insertion of highly resistive contact layers can substantially increase the reader resistance between the top and bottom shields, leading to an increase in the Johnson noise and amplifier noises. Thus, the reader SNR is degraded. As an alternative approach, we considered novel structures at the interfaces between the SL and shields and of the side shields.

Figure 7(a) shows the proposed structure with contact layers with pinholes inserted between the SL and top/bottom shields. The pinholes were made of conductive materials, and the contact layers outside the pinholes were insulating. The pinhole structure suppresses the shunting of the bias current through the top and bottom shields. We simulated $V_H$ of the AHE readers with pinhole contact layers of various pinhole diameters. The pinhole material was assumed to be a 1 nm-thick nonmagnet with $\rho = 25$ $\mu\Omega$ cm (e.g., Ru). The contact layers between SL and side shields were assumed to be a 1 nm-thick nonmagnet without a pinhole structure. Figure 7(b) shows $V_H$ vs. the pinhole diameter for the reader dimension for AD = 4 Tbit/in$^2$ ($t_{SL}$ = 13 nm and $SH = W = 9$ nm). $V_H$ was normalized to the value with uniform conductive contact layers without pinhole structures. $V_H$ increased for smaller pinhole diameters because the shunting of the bias current was more effectively suppressed. We also simulated the case with pinhole contact layers for the four interfaces between the SL and the shields, as shown in Fig. 7(c). The use of the pinhole contact layers between the SL and side shields could also suppress the leakage of $V_H$. Figure 7(d) shows $V_H$ vs. the pinhole diameter for $t_{SL}$ = 13 nm, $SH = W = 9$ nm, $J_x = 1 \times 10^8$ A/cm$^2$, and $\rho_{xy} = 20$ $\mu\Omega$ cm. For a pinhole diameter of 1 nm, $V_H$ reached 2.53 mV, which was close to the value calculated using Eq. (4) of 2.60 mV. Thus, the reduced $V_H$ was almost completely recovered using the pinhole contact layers for the four interfaces.

Figure 7(e) shows another design to suppress the leakage of $V_H$, where the side shields are composed of a laminated structure of soft magnetic layers and highly resistive layers. Such a laminated structure maintains a good electrical conductivity in the film plane ($x$ direction) but suppresses the electrical conductivity in the out-of-plane $z$ direction. Figure 7(f) shows the normalized $V_H$ vs. the resistivity of the 1 nm-thick highly resistive layer ($\rho$). $V_H$ increased with increasing $\rho$ and was saturated at $\rho \sim 1\times10^4$ $\mu\Omega$ cm. The $RA$ value of a 1 nm-thick layer with $\rho$ = $1\times10^4$ $\mu\Omega$ cm is 0.1 $\Omega$ $\mu m^2$, which can be achieved by an MgO tunnel barrier with thickness less than 1 nm. Ferromagnetic layers separated by a tunnel barrier with such an ultralow $RA$ typically exhibit strong ferromagnetic interlayer coupling, which is suitable for aligning the magnetic easy axis of the soft magnetic layers in the $x$ direction to obtain the function of the side shield.

These simulation studies demonstrate that pinhole contact layers and the side shields with a laminated structure of highly resistive layers can effectively suppress the output voltage reduction problem by the bias current shunting and $V_H$ leakage through the shields. The key factor in realizing these novel reader structures is the development of nanofabrication processes, including design, lithography, etching, and deposition.

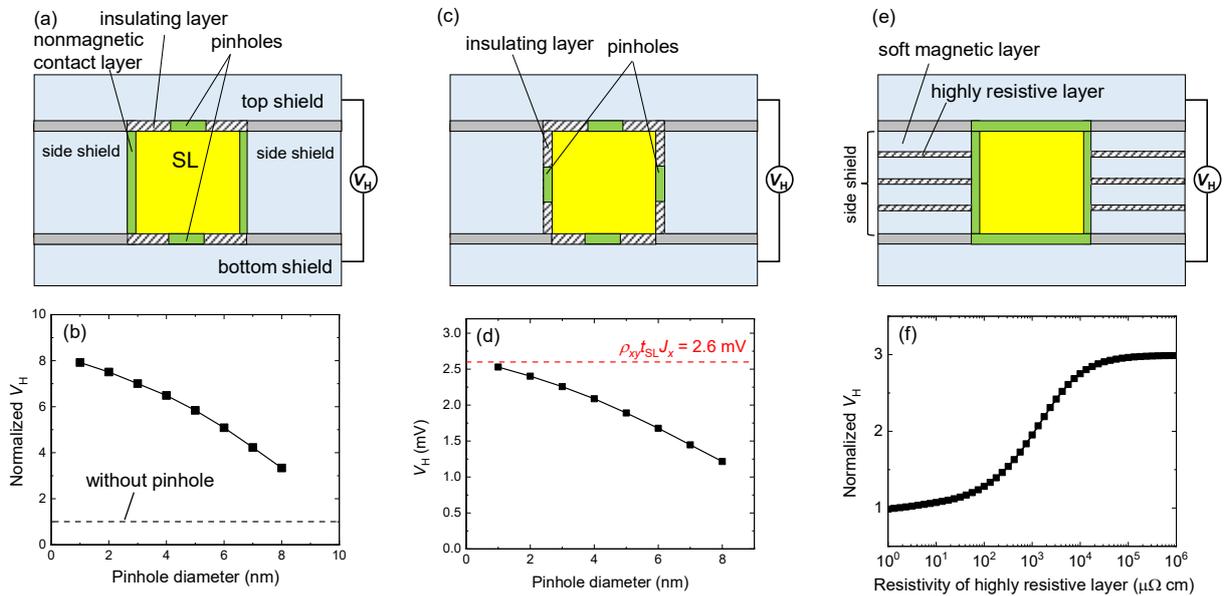

**FIG. 7.** Reader structures used to suppress the output voltage reduction by the shunting of the bias current and/or the leakage of $V_H$. (a) Pinhole contact layers between SL and top/bottom shields, (c) pinhole contact layers at all the interfaces between the SL and shields, and (e) side shields with a laminated structure of soft magnetic layers and highly resistive layers. (b) and

(d) $V_H$ vs. pinhole diameter for the structures of (a) and (c), respectively. (f) $V_H$ vs. the resistivity of 1 nm-thick highly resistive layer laminated in the side shields in (e). In (b) and (f), $V_H$ is normalized by the value without the pinhole contact layer and limited side shield, respectively. For all cases, the bias current is applied between the side shields.

**VII. Summary and perspectives**

The read sensor (reader) technology is one of the major challenges in the realization of HDDs with ultrahigh ADs of over 2 Tbit/in$^2$. Achieving a sufficient SNR in a sensor with a dimension of below 20 nm is considered to be difficult for the current TMR technology because of the multilayered spin-valve structure and the thermal fluctuation of the magnetization of the thin FL. In this paper, we described a simulation-based investigation of magnetic field sensors using the AHE for HDD reader applications. Because an AHE sensor is essentially composed of a single ferromagnetic material, it is a strong candidate for the future reader technology in small dimensions. Although the magnitude of the output voltage by the AHE in conventional ferromagnetic materials is too low for reader applications, the large AHE obtained in topological ferromagnetic materials, such as Co$_2$MnAl and Co$_2$MnGa, has significant potential to exceed the performance of current TMR technology.

Micromagnetic simulations revealed that the AHE reader functions without a lateral bias magnetic field because the shape magnetic anisotropy of the SL spontaneously stabilizes the SL magnetization in the perpendicular direction. The simple relationship that the Hall voltage ($V_H$) is proportional to the orthogonal component of magnetization, AHE readers show $V$–$H_z$ transfer curves with low nonlinearity and zero asymmetry, which is not the case for TMR readers. This feature of the AHE reader can help enhance the SNR by applying a higher reader utilization than that of the conventional TMR reader. The analytical estimation of the reader noise showed that the noise in the TMR reader is dominated by the thermal magnetic noise (mag-noise) because of the small magnetic volume of the thin FL, by which neither an enhancement of the TMR ratio nor a reduction in *RA* contribute to improving the SNR. In contrast, for AHE readers, the contribution of mag-noise is comparable to those of the Johnson noise and amplifier noise. Thus, an enhancement in the anomalous Hall resistivity results in an improvement in the SNR. We also point out that the major challenge for the AHE reader is the reduction in the output voltage due to both shunting of the bias current and leakage of $V_\mathrm{H}$ through the soft magnetic shields surrounding the SL. This problem can be effectively mitigated by a novel reader structure with conductive pinholes in the contact layers inserted between the SL and shields as well as by the side shields composed of a laminated structure with highly resistive layers.

Based on the above investigation, we believe that the following research directions are important for realizing nanoscale HDD readers using AHE.

1) Discovery of new materials with large AHE in thin films. Although the pinhole contact layers and the limited side shields would solve the output voltage reduction problem due to the bias current shunting and $V_\mathrm{H}$ leakage, as shown in Fig. 7, the nanofabrication of such structures is a technological challenge. Therefore, a higher AHE mitigates the requirements for new reader designs and nanofabrication. Obtaining $\rho_{xy}$ values that are substantially greater than 20 $\mu\Omega$ cm should be the target.

Considering that practical HDD readers must be composed of polycrystalline films processed at limited annealing temperatures, typically up to 300 °C, a perfect ordered structure, e.g., the L2$_1$ phase in the case of Heusler alloys, cannot be obtained. Therefore, the prediction of AHE in partially disordered structures, that is, the B2 phase for Heusler alloys, is important for reader applications. From a theoretical viewpoint, the Korringa–Kohn–Rostoker (KKR) method[89,90] with the coherent potential approximation[91] is appropriate for calculating the electronic states of such disordered structures. However, the calculation program for the anomalous Hall conductivity has not been implemented in widely used KKR codes, such as AkaiKKR[92] and SPR-KKR,[93] which is left for future work. Another direction for future theoretical work would be the search for new materials and compositional optimization using first-principles calculations in combination with high-throughput calculations or machine learning, as recently reported in the field of spintronics.[32,94–97]

2) Understanding the device physics of the AHE reader. One of the promising aspects of the AHE reader is that a reader utilization greater than the conventional value of ~0.3 may be applied because of the linear and symmetric $V_H$–$H_z$ transfer curve of AHE devices. This can help enhance the output voltage and SNR of the AHE reader. However, typically, as the swing angle of the SL magnetization increases with increasing reader utilization, the reader operation is destabilized. In particular, when the SL magnetization shows precession due to the ferromagnetic resonance caused by the GHz AC magnetic fields from the recording media, the magnetization response speed to the media field may be degraded. Simulations and experimental studies must be conducted on AHE readers to understand the physics of these spin dynamics. In addition, since AHE readers operate under high bias current densities, it is important to understand the effect of Joule heating on the device performance and reliability. The local Joule heating of the AHE SL and the heat dissipation to the surrounding NiFe shields can cause thermoelectric and spin-caloritronic effects, such as Seebeck effect and anomalous Nernst effect, which is of another fundamental and practical interest.

3) Novel reader structure design and nanofabrication process development. Suppressing the output voltage reduction by bias current shunting and $V_H$ leakage is a major technological challenge for realizing AHE readers. To achieve this, the design of the reader structure and the development of the nanofabrication process are critical. In particular, aligning the top and bottom contact layers with pinholes in nanometer-scale position precision is a significant challenge.

The above efforts require collaboration of researchers and engineers with various expertise, including condensed matter physics, materials science, device physics, and process technology. We anticipate joint efforts from academia and industry to realize AHE readers and other magnetic sensing devices using the AHE, which will support our future society.

**Acknowledgements**

We acknowledge Dr. Thomas Coughlin for providing us with an analysis of storage demand and reference information. We thank Keigo N. for his help with the illustrations. This work was supported by Advanced Storage Research Consortium (ASRC), JST-CREST (Grant No. JPMJCR21O1), and the MEXT Program: Data Creation and Utilization-Type Materials Research and Development Project (Digital Transformation Initiative Center for Magnetic Materials; Grant No. JPMXP1122715503), Japan.

**Data Availability Statement**

The data supporting the findings of this study are available from the corresponding author upon reasonable request.

**Conflict of interest**

The authors have no conflict of interest regarding the publication of this article.

**Appendix: SNR calculations**

The output voltage of the AHE reader can be calculated using the expression:

$$\Delta V_{\text{AHE}} = 2\rho_{xy} J_x t_{\text{SL}} \eta_{\text{AHE}}, \qquad (A1)$$

where $\eta_{\text{AHE}}$ is the reader utilization. The output voltage of the TMR reader can be calculated as follows:

$$\Delta V_{\text{TMR}} = \Delta R/R \cdot V_b \eta_{\text{TMR}}, \qquad (A2)$$

where $\Delta R/R$ is the TMR ratio. The reader utilization of both types of readers are not necessarily the same. Indeed, $\eta_{\text{AHE}}$ may be larger than $\eta_{\text{TMR}}$ owing to the higher linearity and the perfect symmetry of the $V_H$–$H_z$ curve of the AHE readers (Figs. 4(f) and (g)). However, for simplicity, we assumed the same reader utilization values: $\eta_{\text{AHE}} = \eta_{\text{TMR}} = 0.3$.

The noise voltage densities in the unit of V/$\sqrt{\text{Hz}}$ were calculated using the following theoretical expressions:

Johnson noise (AHE): $\quad \sqrt{4k_B T R} \qquad (A3)$

Shot noise (TMR): $\quad \sqrt{2eV_b R \coth(eV_b/2k_B T)} \qquad (A4)$

Mag-noise: $\quad \Delta V_{\max} \cdot \sqrt{\dfrac{k_B T \alpha}{\gamma (H_{\text{stiff}})^2 M_s V_{\text{FL}}}} \qquad (A5)$

Amplifier noise: $\quad \sqrt{V_n^2 + (I_n R)^2} \qquad (A6)$

where $k_B$: Boltzmann constant, $T$: temperature, $R$: reader resistance, $e$: elementary charge, $\alpha$: magnetic damping constant of SL and FL, $\gamma$: gyromagnetic ratio of SL and FL, $H_{\text{stiff}}$: magnetic bias (stiffness) field, $M_s$: saturation magnetization, $V_{\text{FL}}$: volume of the SL and FL, $V_n$: voltage noise of the amplifier, and $I_n$: current noise of the amplifier. $\Delta V_{\max}$ is the output voltage when $\eta = 1$ in Eqs. (A1) and (A2). The stripe height (*SH*) was assumed to be $SH = 1.2W$, where $W$ is the reader width. For both the AHE and TMR readers, the following variables were fixed: *T*

= 350 K, $\alpha$ = 0.02, $\gamma$ = 1.76 × 10$^{11}$ rad s$^{-1}$ T$^{-1}$, and $M_s$ = 1000 emu/cm$^3$. Although the SL magnetization of the AHE reader may be spontaneously stabilized by the shape anisotropy and its anisotropy depends on the SL dimension as discussed in the main text and in Fig. 4, for simplicity, we assumed $H_{\text{stiff}}$ = 800 Oe for both the AHE and TMR readers. We adopted amplifier noise with $V_n$ = 1.2 nV/$\sqrt{\text{Hz}}$ and $I_n$ = 6 pA/$\sqrt{\text{Hz}}$ from Ref.[66]. From Eq. (A6), the amplifier noise is higher for higher $R$. Thus, the improvement of the amplifier with lower current noise is another critical task. The total noise was calculated from the square root of the sum of the square of each noise (Eqs. (A3)–(A6)), i.e.,

$$N_{\text{total}} = \sqrt{N_{\text{Johnson}}^2 + N_{\text{mag}}^2 + N_{\text{amp}}^2} \quad \text{(A7) for AHE reader, and}$$

$$N_{\text{total}} = \sqrt{N_{\text{shot}}^2 + N_{\text{mag}}^2 + N_{\text{amp}}^2} \quad \text{(A8) for TMR reader.}$$

The noise voltage was calculated by multiplying the bandwidth ($\Delta f$), which corresponds to the maximum frequency of the AC magnetic field from the recording bits and calculated by the linear density of the recording bit and the disk rotation speed. Table A1 shows the linear densities (in kilo flux changes per inch, kFCI) for ADs in the range of 2.4–6.0 Tbit/in$^2$ predicted by Albuquerque et al.[66] We calculated $\Delta f$ at the outer diameter of a 3.5-inch disk rotating at 7200 rpm.

**Table A1.** Linear densities of the recording bit and the recording bandwidths ($\Delta f$) for ADs in the ranges of 2.4−6.0 Tbit/in$^2$.

| AD (Tbit/in$^2$) | Linear density (kFCI) | $\Delta f$ (GHz) |
|---|---|---|
| 2.4 | 2700 | 1.78 |
| 4.0 | 3400 | 2.24 |
| 6.0 | 3900 | 2.57 |